\documentstyle[12pt,epsf]{article}
\def\href#1#2{#2}   
\newif\ifdraft
\draftfalse	  
\reversemarginpar   
\makeatletter
\let\mlabel=\label
\let\adkendequation=\endequation%
\def\endequation{\adkendequation\adklabel\global\@ignoretrue}
\let\adkendeqnarray=\endeqnarray%
\def\endeqnarray{\adkendeqnarray\adklabel\global\@ignoretrue}
\newbox\marglabbox
\def\adklabel{\ifvoid\marglabbox\else\marginpar{\unhbox\marglabbox}\fi}
\def\label#1{\ifdraft\ifmmode%
  \global\setbox\marglabbox=\hbox{\hfill\fbox{\tiny\verb*~#1~}}%
  \else\ifinner\else\marginpar{\hfill\fbox{\tiny\verb*~#1~}}%
  \fi\fi\fi \mlabel{#1}}
\makeatother
\ifdraft%
\fi
\font\twelvebb=msbm12
\font\tenbb=msbm10
\font\sevenbb=msbm7
  \newfam\bbfam
  \def\bb{\fam\bbfam\twelvebb}
  \textfont\bbfam=\twelvebb
  \scriptfont\bbfam=\tenbb
  \scriptscriptfont\bbfam=\sevenbb
\font\twelveeusm=eusm10 scaled 1200
\font\teneusm=eusm10
  \newfam\eusmfam
  
  \textfont\eusmfam=\twelveeusm
  \scriptfont\eusmfam=\teneusm
  \scriptscriptfont\eusmfam=\scriptfont\eusmfam
\font\twelvefrak=eufm10 scaled 1200
\font\tenfrak=eufm10
  \newfam\frakfam
  
  \textfont\frakfam=\twelvefrak
  \scriptfont\frakfam=\tenfrak
  \scriptscriptfont\frakfam=\scriptfont\frakfam
\def\sqr#1#2{{\vcenter{\hrule height.#2pt
   \hbox{\vrule width.#2pt height#1pt \kern#1pt
      \vrule width.#2pt}
   \hrule height.#2pt}}}

\def\bsqr#1#2{{\vrule width #1pt height#2pt}}
\def\bsquare{{\mathchoice\bsqr66\bsqr66\bsqr33\bsqr33}}
\def\badbreak{\penalty1000}

\def\identity{{\bb I}}			    

\def\rational#1#2{{\mathchoice{\textstyle{#1\over#2}}%
  {\scriptstyle{#1\over#2}}{\scriptscriptstyle{#1\over#2}}{#1/#2}}}
\def\half{\rational12}			    
\newcommand{\D}{{\bf D}}                    
\newcommand{\Rb}{{\bf R}}                   
\newcommand{\gone}{\gamma_{1}}              
\newcommand{\gtwo}{\gamma_{2}}              
\newcommand{\gfive}{\gamma_{5}}             

\begin{document}

\title{
  Comment on {\tt hep-lat/9901005 v1-v3} by W.~Bietenholz}

\author{Ivan Horv\'ath\thanks{{\tt ih3p@virginia.edu}} \\[1ex]
  Department of Physics, University of Virginia\\
  Charlottesville, Virginia 22903, U.S.A.}

\date{March 12, 2000}

\maketitle

\begin{abstract}
  \noindent
  I comment on the above paper(s) discussing the issue of non-ultralocality
  for Ginsparg-Wilson fermionic actions. The purpose of this note is to point 
  out that the new claim in ``v3'' (Feb 24, 2000), alleging the proof of 
  non-ultralocality for ``all Ginsparg-Wilson fermions'', might not be 
  substantiated. The remarkable evolution of versions of this paper is put in 
  context with the literature existing at the time of their appearance. 
\end{abstract}
                   
\section{The Setting}

One of the interesting issues in lattice field theory is understanding the 
structural properties of Ginsparg-Wilson (GW) fermionic actions, i.e. the 
actions for which the chirally nonsymmetric part of the massless propagator 
$\Rb \equiv (\D^{-1})_N \ne 0$ is local \footnote{For definitions, see Ref.~[2]}. 
Perhaps the most basic problem is clarifying under which conditions the GW 
property is (in)compatible with ultralocality of the action. This question can 
be usefully discussed already at the free level, because non-ultralocality of 
a free action implies non-ultralocality in any gauge-invariant interacting 
theory based on it.
 
It turns out that a lot can be said if one requires the underlying free theory
to respect the symmetries of the hypercubic lattice structure~[1,2], i.e.
translations and transformations of the hypercubic group. In fact, this is
naturally the most important case since ideally, one prefers to work with
the action respecting all crucial fundamental symmetries, i.e. hypercubic 
symmetries, gauge invariance, and chiral symmetry (in this case 
Ginsparg-Wilson-L\"uscher (GWL) symmetry). It is worth noting that attempts to 
clarify this issue in the context where symmetry under hypercubic group is 
abandoned and only lattice translations are kept, were not very successful so far.

Current insight into the question of (in)compatibility of hypercubic symmetries 
with ultralocality of GW actions is based on two suggestions: 

\noindent\underline{{\bf Part 1:}} Studying the consequences of the GW property 
   for ultralocal lattice Dirac operators restricted on lines corresponding to 
   periodic directions in the Brillouin zone. This was introduced in Ref.~[1] 
   and fully developed in Ref.~[2].
\medskip

\noindent\underline{{\bf Part 2:}} Studying the consequences of the GW property 
   at the origin of the Brillouin zone for two-dimensional restrictions of 
   ultralocal lattice Dirac operators. This was introduced in Ref.~[3] with the 
   suggestion that corresponding analytic properties at the origin have powerful 
   global consequences for the operator in ultralocal case, and may lead to the 
   necessity of fermion doubling.

\section{The Literature}

It is useful to outline the evolution of the above issues in the literature
chronologically. For clarity, I will refer to various versions of paper 
{\tt hep-lat/9901005} [4-6] as ``v1-v3''.
\medskip

{\bf (1)}$\;$ In Ref.~[1], the idea of \underline{Part 1} was introduced and
   necessary steps were performed to prove that canonical GW operators, i.e. 
   operators satisfying 
   $\{\D,\gfive \} \,=\, \D \gfive \D $, or 
   $\Rb\equiv (\D^{-1})_N = \half\identity$, can not be ultralocal. On the first 
   page of that Letter it is also explicitly stated that the proof can be extended 
   to all ultralocal $\Rb$, trivial in Dirac space. This and other simple 
   generalizations were explicitly deferred to Ref.~[2] for reasons of space.
\medskip

{\bf (2)}$\;$ The paper ``v1'' adopts the approach of \underline{Part 1}, and the 
   claim is made in the abstract of extending the proof of Ref.~[1] to a
   ``...much larger class of Ginsparg-Wilson fermions...''. While it is quite 
   unclear from the paper what this larger class is, it is explicitly stated
   that the alleged proof applies to all cases for which $\Rb^{-1}$ is ultralocal,
   where $\Rb$ is a ``Dirac scalar'' i.e. trivial in spinor space 
   (Ref.~[4], pages 1,2). This would be an interesting new result but, 
   unfortunately, it was not substantiated.
\medskip

{\bf (3)}$\;$ In paper ``v2'' the above claim of ``v1'' is changed, and it is  
   stated that the proof rather applies to all cases for which $\Rb$ is ultralocal 
   and trivial in spinor space (Ref.~[5], pages 1,2). This is a result put 
   forward in Ref.~[1].
\medskip  

\noindent While ``v2'' uses the ingredients of \underline{Part 1},
   the satisfactory discussion of its merits as a proof (and the merits of its 
   assumptions) would be rather involved. Some improvements were put
   forward later in ``v3'', and this latest version will be discussed
   in Sec.~3 of this Comment.
\vskip 0.08in

{\bf (4)} Paper [2] describes in detail the consequences of the approach 
   of \underline{Part 1}, as indicated in Ref.~[1]. The considerations on periodic 
   directions of the Brillouin zone are shown to be sufficient to prove that 
   infinitesimal GWL symmetry transformations must be non-ultralocal for arbitrary 
   GW action in the presence of hypercubic symmetries 
   {\it (``weak non-ultralocality'')}. On the basis of {\it weak 
   non-ultralocality} it is then proved that GW operators for which 
   $\Rb\equiv (\D^{-1})_N$ is ultralocal, can not be ultralocal. This
   contains the result announced in Ref.~[1], and generalizes it
   further by showing that triviality in spinor space is not crucial. 
\medskip 

{\bf (5)} The approach of \underline{Part 2} is proposed in 
   Ref.~[3] \footnote{Ref.~[3] is a contribution of the author to the proceedings 
   of the conference ``Lattice Fermions and the Structure of the Vacuum'', Dubna, 
   Russia, Oct 5-9 1999. The list of participants at this conference includes the 
   author of [4-6].}. 
   It is pointed out that in the presence of hypercubic symmetries, 
   the GW condition for ultralocal actions translates into analytic properties of 
   certain {\it rational functions} after suitable change of variables, and  
   that two-dimensional restrictions of operators on the Brillouin zone 
   already capture the required analytic structure. The crucial observation 
   is that analyticity at the origin implies factorization properties of involved 
   polynomials, which strongly constraints the global behaviour of the 
   corresponding action, and may imply fermion doubling. This connection is 
   encapsulated in the hypothesis (Ref.~[3], page 6), reflecting the conjectured 
   property of such factorizations. The hypothesis was proposed as a key to the 
   problem of {\it ``strong non-ultralocality''}, i.e. non-ultralocality of all 
   doubler-free GW actions in the presence of hypercubic symmetries, as formulated 
   in Ref.~[2]. There is no resolution of the hypothesis to date.
\medskip

{\bf (6)} The version ``v3'' appears (Feb 24, 2000) with two major changes compared 
          to ``v2'':
       \begin{description}

         \item{$(\alpha)$} Parts of formalism and the main claim of ``v2'' are 
         upgraded to the most general result of Ref.~[2], i.e. discussion involves
         actions with  $\Rb \equiv (\D^{-1})_N$ ultralocal (not just trivial in 
         spinor space). This now forms the STEP 1 of the paper. 

         \item{$(\beta)$} The approach \underline{Part 2} of Ref.~[3] is adopted 
         in the completely new part STEP 2. The argument is presented in such
         a way that STEP 1 and STEP 2 together are claimed to imply the
         non-ultralocality for ``all Ginsparg-Wilson fermions''~[6].

      \end{description}

\section{Discussion of ``v3''}

    The purpose of this note is to point out that one can raise several objections 
    to the arguments of ``v3'', casting doubts about the result claimed in the 
    paper. Some of these objections are described below.
    \medskip

{\bf (a)} One of the starting points of ``v3'' is the suggestion that Eq.~(2) 
      represents the general ansatz for the restriction of arbitrary lattice 
      Dirac operator in $d$ dimensions to the two-dimensional momentum plane through 
      $\D \rightarrow \D(p_1,p_2,0,\ldots,0)$. This is supposed to be true if ``We 
      assume Hermiticity, discrete translation invariance, as well as invariance under 
      reflections and exchange of the axes.''~[6]\footnote{By ``Hermiticity'', the
      author of ``v3'' perhaps means $\gfive$-Hermiticity. However, there doesn't 
      appear to be a good reason or necessity to assume either of these 
      (none is assumed Refs~[1-3]).}.
      If true, the statement of this nature should perhaps be proved. If not true,
      the additional possible terms (such as one proportional to $\gone\gtwo$,
      which is compatible with hypercubic symmetries) must be included and
      the proof should proceed with the presence of such terms.
      \medskip

{\bf (b)} The GW property is encoded in the analyticity properties of 
      Clifford components of $(\D^{-1})_N$. ``v3'' uses the non-invertible 
      change of variables (two to one) on the Brillouin zone 
      $c_\mu = 1 - \cos p_\mu$ (see Eq.~(14)),
      and implicitly assumes that the required analyticity properties are
      inherited in new variables (see case $(b)$ on page 6 of ``v3'').
      In the proof one would expect the required analyticity properties 
      to be defined, as well as careful justification that the above change
      of variables preserves them.
      \medskip

{\bf (c)} Eq.~(18) of ``v3'' introduces the following polynomial decomposition 
      for symmetric polynomial $K(c_1,c_2)$ (the fact that $K$ must be symmetric
      has not been justified) 
      \begin{displaymath}
         K(c_1,c_2) \;=\; c_1 X(c_1,c_2) \,+\, c_2 X(c_2,c_1)
      \end{displaymath}
      Since $K(0,0)=0$, the required polynomial $X$ exists, but this representation 
      is not unique. There are infinitely many $X$ that represent the same $K$, 
      for example, $X\; \rightarrow\; X + c_2(c_1-c_2)$. How then is $X$ fixed?
      \medskip
 
{\bf (d)} The crucial part of the argument presented in STEP 2 of ``v3'' 
      is the identification expressed by Eq.~(22), which is supposed to follow
      from Eqs.~(16,17) and (21). Unless there are hidden assumptions
      and arguments, there doesn't appear to be any reason why this 
      identification should hold. If some symmetric polynomial $P(c_1,c_2)$
      can be written in terms of another polynomial $Q(c_1,c_2)$ as
      \begin{displaymath}
         P(c_1,c_2) \;=\; c_1(2-c_1) Q(c_1,c_2) \,+\, c_2(2-c_2) Q(c_2,c_1)
      \end{displaymath}
      then for similar reasons to those discussed in item (c) above, there
      are infinitely many polynomials $Q$ that can be used for this
      decomposition. Since there is no uniqueness, how does the Eq.~(22)
      follow?\footnote{It also appears that the generic case
      $n_1=n_2=0$ of ``v3'' should be discussed separately, because the
      polynomial structure is different.} 
      If a conclusion of such nature is possibly justifiable, then the 
      proof would seem to require an explicit argument to that effect.
      \medskip      
     
\noindent It should be emphasized in closing that the above objections are not 
      aimed at excessive improvements of rigor in ``v3''. The aim is to point
      out that there appear to be serious holes in the arguments, raising the
      worry that the claim of ``v3'' might simply not be justified at all. 
      The problem of {\it ``strong non-ultralocality''} of GW fermions is 
      an important issue and it would be inherently useful to resolve it cleanly 
      (by either giving a proof or a counterexample). Hopefully, the remarks in 
      this Comment can contribute to eventually achieving that goal.
      \medskip    

{\bf Acknowledgement:} I thank R.~Mendris for many pleasant discussions
on the issues related to those discussed here.

\end{document}
\bye